\documentclass[iop]{emulateapj}
\usepackage{amsfonts}
\usepackage{rotating}
\usepackage{multirow}

\def \etal {{\em et~al.}\ $\!\!$} 
\def \note #1 {{\tt [NOTE: #1] }}

\def \nustar {{\em NuSTAR\ }}
\def \swift {{\em Swift\ }}

\def \swiftbat {{\em Swift}/BAT\ }

\def \suzaku {{\em Suzaku\ }}

\def \integral {{\em INTEGRAL\ }}
\def \bepposax {{\em BeppoSAX\ }}

\def \xspec {{\tt Xspec\ }}
\def \pexrav {{\tt pexrav\ }}
\def \pexmon {{\tt pexmon\ }}

\def \mcg {{MCG\,--05-23-016\ }}

\slugcomment{  }
\shorttitle{Coronal Properties of \mcg with {\em NuSTAR}}
\shortauthors{Balokovi\'{c}, M. {\em et~al.}}

\begin{document}

\title{Coronal Properties of the Seyfert~1.9 Galaxy \mcg\\Determined from Hard X-ray Spectroscopy with \nustar}

\author{
M.~Balokovi\'{c}\altaffilmark{1},
G.~Matt\altaffilmark{2},
F.\,A.~Harrison\altaffilmark{1},
A.~Zoghbi\altaffilmark{3,4},
D.\,R.~Ballantyne\altaffilmark{5},
S.\,E.~Boggs\altaffilmark{6},
F.\,E.~Christensen\altaffilmark{7},
W.\,W.~Craig\altaffilmark{6,8},
C.\,J.~Esmerian\altaffilmark{1},
A.\,C.~Fabian\altaffilmark{9},
F.~F{\" u}rst\altaffilmark{1},
C.\,J.~Hailey\altaffilmark{10},
A.~Marinucci\altaffilmark{2},
M.\,L.~Parker\altaffilmark{9},
C.\,S.~Reynolds\altaffilmark{3,4},
D.~Stern\altaffilmark{11}
D.\,J.~Walton\altaffilmark{1},
W.\,W.~Zhang\altaffilmark{12}
}

\altaffiltext{1}{Cahill Center for Astronomy and Astrophysics, Caltech, Pasadena, CA 91125, USA}
\altaffiltext{2}{Dipartimento di Matematica e Fisica, Universit{\` a} degli Studi Roma Tre, via della Vasca Navale 84, I-00146 Roma, Italy}
\altaffiltext{3}{Department of Astronomy, University of Maryland, College Park, MD 20742-2421, USA}
\altaffiltext{4}{Joint Space-Science Institute (JSI), College Park, MD 20742-2421, USA}
\altaffiltext{5}{Center for Relativistic Astrophysics, School of Physics, Georgia Institute of Technology, Atlanta, GA 30332}
\altaffiltext{6}{Space Science Laboratory, University of California, Berkeley, CA 94720, USA}
\altaffiltext{7}{DTU Space National Space Institute, Technical University of Denmark, Elektrovej 327, 2800 Lyngby, Denmark}
\altaffiltext{8}{Lawrence Livermore National Laboratory, Livermore, CA 94550, USA}
\altaffiltext{9}{Institute of Astronomy, Madingley Road, Cambridge CB3 0HA, UK}
\altaffiltext{10}{Columbia Astrophysics Laboratory, Columbia University, New York, NY 10027, USA}
\altaffiltext{11}{Jet Propulsion Laboratory, California Institute of Technology, Pasadena, CA 91109, USA}
\altaffiltext{12}{NASA Goddard Space Flight Center, Greenbelt, MD 20771, USA}

%%%%%%%%%%%%%%%%%%%%%%%%%%%%%%%%%%%%%%%%%%%%%%%%%%%%%%%%%%%%%%%%%%%%%%%%%%%%%%%%%%%%%%%%%%%%%%%%%%
\begin{abstract} %%%%%%%%%%%%%%%%%%%%%%%%%%%%%%%%%%%%%%%%%%%%%%%%%%%%%%%%%%%%%%%%%%%%%%%%%%%%%%%%%

Measurements of the high-energy cut-off in the coronal continuum of active galactic nuclei have long been elusive for all but a small number of the brightest examples. We present a direct measurement of the cut-off energy in the nuclear continuum of the nearby Seyfert~1.9 galaxy \mcg with unprecedented precision. The high sensitivity of \nustar up to 79~keV allows us to clearly disentangle the spectral curvature of the primary continuum from that of its reflection component. Using a simple phenomenological model for the hard X-ray spectrum, we constrain the cut-off energy to $116_{-5}^{+6}$~keV with 90\% confidence. Testing for more complex models and nuisance parameters that could potentially influence the measurement, we find that the cut-off is detected robustly. We further use simple Comptonized plasma models to provide independent constraints for both the kinetic temperature of the electrons in the corona and its optical depth. At the 90\% confidence level, we find $kT_e=29\pm2$~keV and $\tau_e=1.23\pm0.08$ assuming a slab (disk-like) geometry, and $kT_e=25\pm2$~keV and $\tau_e=3.5\pm0.2$ assuming a spherical geometry. Both geometries are found to fit the data equally well and their two principal physical parameters are correlated in both cases. With the optical depth in the $\tau_e\gtrsim1$ regime, the data are pushing the currently available theoretical models of the Comptonized plasma to the limits of their validity. Since the spectral features and variability arising from the inner accretion disk have been observed previously in \mcg$\!\!$, the inferred high optical depth implies that a spherical or disk-like corona cannot be homogeneous.

\end{abstract} %%%%%%%%%%%%%%%%%%%%%%%%%%%%%%%%%%%%%%%%%%%%%%%%%%%%%%%%%%%%%%%%%%%%%%%%%%%%%%%%%%%
%%%%%%%%%%%%%%%%%%%%%%%%%%%%%%%%%%%%%%%%%%%%%%%%%%%%%%%%%%%%%%%%%%%%%%%%%%%%%%%%%%%%%%%%%%%%%%%%%%


%%%%%%%%%%%%%%%%%%%%%%%%%%%%%%%%%%%%%%%%%%%%%%%%%%%%%%%%%%%%%%%%%%%%%%%%%%%%%%%%%%%%%%%%%%%%%%%%%%
%%%%%%%%%%%%%%%%%%%%%%%%%%%%%%%%%%%%%%%%%%%%%%%%%%%%%%%%%%%%%%%%%%%%%%%%%%%%%%%%%%%%%%%%%%%%%%%%%%
\section{Introduction} %%%%%%%%%%%%%%%%%%%%%%%%%%%%%%%%%%%%%%%%%%%%%%%%%%%%%%%%%%%%%%%%%%%%%%%%%%%
%%%%%%%%%%%%%%%%%%%%%%%%%%%%%%%%%%%%%%%%%%%%%%%%%%%%%%%%%%%%%%%%%%%%%%%%%%%%%%%%%%%%%%%%%%%%%%%%%%
%%%%%%%%%%%%%%%%%%%%%%%%%%%%%%%%%%%%%%%%%%%%%%%%%%%%%%%%%%%%%%%%%%%%%%%%%%%%%%%%%%%%%%%%%%%%%%%%%%

\label{sec:intro}

The intrinsic X-ray continuum of active galactic nuclei (AGN) is thought to be produced in the immediate vicinity of the central black hole. Phenomenologically, the nuclear continuum can be described as a power law, typically with a photon index of 1.8--2.0, with an exponential cut-off at 150--350~keV \citep{dadina-2007,burlon+2011,molina+2013,vasudevan+2013,malizia+2014,ballantyne-2014}. The currently accepted model for formation of this spectral component is the inverse Compton scattering of the thermal radiation from the accretion disk by relativistic electrons distributed around the black hole in a structure referred to as the corona (e.g., \citealt{rybicki+lightman-1979,titarchuk-1994,zdziarski+2000}). The shape of the coronal spectrum is a function of the seed photon field, the kinetic temperature of the plasma, the geometry of the corona and the observer orientation. 

Previous studies suggest that the corona does not uniformly cover the surface of the accretion disk \citep{haardt+1994}, and that it is likely compact \citep{reis+miller-2013}. Microlensing measurements on distant quasars confirm the compactness of the X-ray-emitting region (e.g. \citealt{dai+2010,mosquera+2013}). However, other physical parameters of AGN coronae are currently poorly constrained due to the lack of direct observations in the hard X-ray band, as well as the degeneracy introduced by contributions from the processed (reflected) spectra from the inner regions of the accretion disk and the dusty molecular torus at larger distances (e.g., \citealt{george+fabian-1991,ghisellini+1994}). Disentangling those spectral components requires high-quality hard X-ray data.

We report on the high-energy cut-off measurement and coronal parameters of the active nucleus of the nearby ($z=0.0085$; 36~Mpc) Seyfert~1.9 galaxy \mcg \citep{veron+1980,wegner+2003}, using \nustar data in the 3--79~keV band \citep{harrison+2013}. This AGN has been extensively observed in the soft X-ray band \citep{weaver+1997,mattson+weaver-2004,balestra+2004,braito+2007,reeves+2007,zoghbi+2013}, revealing a complex structure of the flourescent line emission, including both broad and narrow components produced by the disk and the torus reflection, respectively. The high-energy spectrum, however, has been only poorly constrained thus far: e.g., \citet{perola+2002} and \citet{dadina-2007} found high-energy cut-offs at $147_{-40}^{+70}$ and $190_{-60}^{+110}$~keV from \bepposax data, \citet{molina+2013} found it at $85_{-20}^{+35}$~keV using \integral data, \citet{beckmann+2008} combined \swiftbat and \integral to support a variable cut-off between 50 and $\gtrsim$100~keV, while other results in the literature only placed lower limits in the 100--200~keV range. The main reason for the discrepant measurements in the past is likely the degeneracy between a cut-off at $\lesssim\!200$~keV and a strong reflection continuum. The high signal-to-noise ratio achieved in the observations of \mcg with \nustar allows us to clearly separate the spectral curvature due to the reflection continuum from the spectral curvature due to the coronal cut-off. In \S\,\ref{sec:observations} we report on the \nustar observations and in \S\,\ref{sec:modeling} we present our spectral analysis. In \S\,\ref{sec:discussion} we discuss the potential issues and the physical properties of the corona, and briefly summarize our results in \S\,\ref{sec:summary}.

%%%%%%%%%%%%%%%%%%%%%%%%%%%%%%%%%%%%%%%%%%%%%%%%%%%%%%%%%%%%%%%%%%%%%%%%%%%%%%%%%%%%%%%%%%%%%%%%%%
%%%%%%%%%%%%%%%%%%%%%%%%%%%%%%%%%%%%%%%%%%%%%%%%%%%%%%%%%%%%%%%%%%%%%%%%%%%%%%%%%%%%%%%%%%%%%%%%%%
\section{Observations and Data} %%%%%%%%%%%%%%%%%%%%%%%%%%%%%%%%%%%%%%%%%%%%%%%%%%%%%%%%%%%%%%%%%%
%%%%%%%%%%%%%%%%%%%%%%%%%%%%%%%%%%%%%%%%%%%%%%%%%%%%%%%%%%%%%%%%%%%%%%%%%%%%%%%%%%%%%%%%%%%%%%%%%%
%%%%%%%%%%%%%%%%%%%%%%%%%%%%%%%%%%%%%%%%%%%%%%%%%%%%%%%%%%%%%%%%%%%%%%%%%%%%%%%%%%%%%%%%%%%%%%%%%%

\label{sec:observations}

\nustar observed \mcg on two occasions: on 2012 July 11--12 (OBSID 10002019), and on 2013 June 3--7 (OBSID 60001046). The first observation was conducted as a part of the \nustar calibration campaign. The second observation was a science observation carried out simultaneously with a long \suzaku observation.  We defer the broadband (0.5--79~keV) spectral analysis of the simultaneous \nustar and \suzaku data taken in 2013 to a forthcoming paper (Zoghbi {\em et al.}, {\em in prep.}). Hereafter, we refer to the 2012 and 2013 observations as the {\it calibration} and {\it science} observations, respectively. The event files were cleaned and processed using the NuSTARDAS software package (version 1.2.1) and the scripts {\tt nupipeline} and {\tt nuproducts} \citep{perri+2013}. After the automated processing by the pipeline, the total source exposure is 34~ks for the calibration observation, and 160~ks for the science observation. We extracted the source spectra from circular regions 120\arcsec\ in radius, centered on the peak of the source image. Corresponding background spectra were extracted from polygonal regions encompassing the same detector, but avoiding the region within 140\arcsec\ from the source image peak. We estimate that at most 2\% of the background counts above 25~keV can be due to contamination by the source. The response matrices were generated using the calibration database (CALDB) version 20131223.

The analysis presented here is based predominantly on the higher-quality science observation, while the calibration observation is used to investigate the spectral variability on the timescale of one year. The count rate was variable at the level of $\lesssim30$\% during the long \nustar science observation, and $\lesssim$20\% during the calibration observation. The variability on timescales of $\lesssim$1-ks is addressed in detail in a separate publication \citep{zoghbi+2014}. For the analysis presented in this paper, we use the observation-averaged spectra from each of the two \nustar focal plane modules (FPMA and FPMB), and fit them jointly for each of the two observations, allowing for the cross-normalization constant to vary freely in all fits. The normalization offset is found to be smaller than 5\% in all cases, as expected from instrument calibration \citep{madsen+2015}.

%%%%%%%%%%%%%%%%%%%%%%%%%%%%%%%%%%%%%%%%%%%%%%%%%%%%%%%%%%%%%%%%%%%%%%%%%%%%%%%%%%%%%%%%%%%%%%%%%%
%%%%%%%%%%%%%%%%%%%%%%%%%%%%%%%%%%%%%%%%%%%%%%%%%%%%%%%%%%%%%%%%%%%%%%%%%%%%%%%%%%%%%%%%%%%%%%%%%%
\section{Spectral Modeling} %%%%%%%%%%%%%%%%%%%%%%%%%%%%%%%%%%%%%%%%%%%%%%%%%%%%%%%%%%%%%%%%%%%%%%
%%%%%%%%%%%%%%%%%%%%%%%%%%%%%%%%%%%%%%%%%%%%%%%%%%%%%%%%%%%%%%%%%%%%%%%%%%%%%%%%%%%%%%%%%%%%%%%%%%
%%%%%%%%%%%%%%%%%%%%%%%%%%%%%%%%%%%%%%%%%%%%%%%%%%%%%%%%%%%%%%%%%%%%%%%%%%%%%%%%%%%%%%%%%%%%%%%%%%

\label{sec:modeling}

We model the \nustar data in \xspec (version 12.8.1; \citealt{arnaud+1996}) using $\chi^2$ statistics. In order for $\chi^2$ statistics to provide unbiased results we group the data to have a signal-to-noise ratio of at least 10 per bin after background subtraction. All uncertainties on spectral parameters are reported as 90\% confidence intervals from marginalized probability distributions determined using the Markov Chain Monte Carlo (MCMC) algorithm available in \xspec$\!\!$.

%%%%%%%%%%%%%%%%%%%%%%%%%%%%%%%%%%%%%%%%%%%%%%%%%%%%%%%%%%%%%%%%%%%%%%%%%%%%%%%%%%%%%%%%%%%%%%%%%%
\begin{figure*}[t!] %%%%%%%%%%%%%%%%%%%%%%%%%%%%%%%%%%%%%%%%%%%%%%%%%%%%%%%%%%%%%%%%%%%%%%%%%%%%%%
\begin{center}
\includegraphics[width=2\columnwidth]{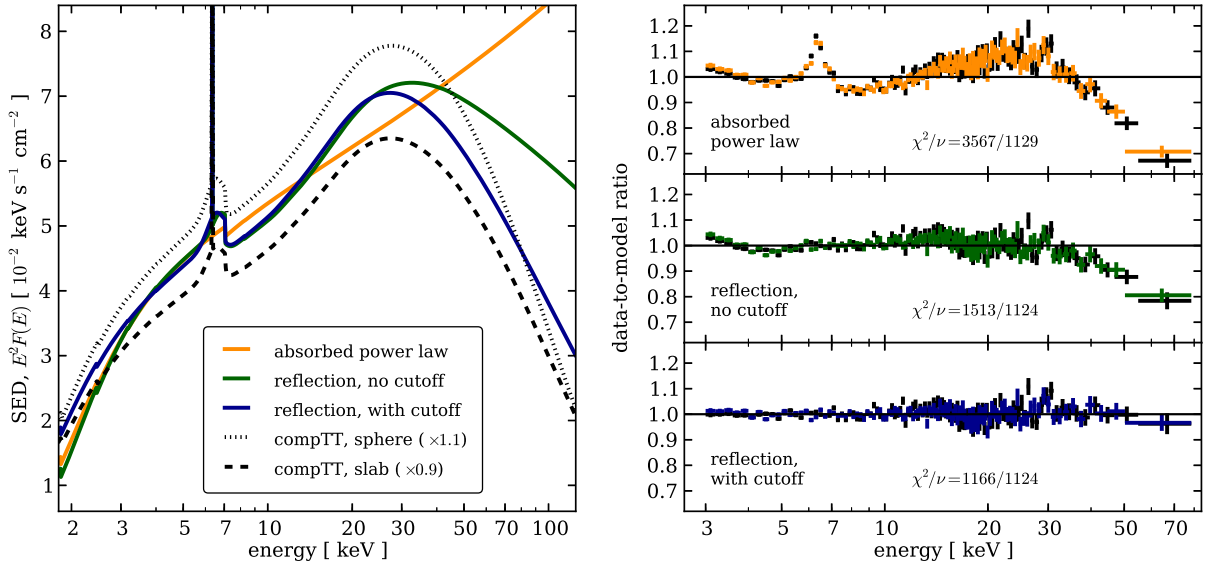}
\caption{ {\it Left:} The model curves for the best-fit models considered in this work: phenomenological ones in solid colored lines and physical ones in dotted and dashed black lines. The physical {\tt compTT} model for spherical (slab) geometry has been moved up (down) by 10\% for clarity. {\it Right:} The data-to-model ratios for the \nustar science observation data, and the three phenomenological models discussed in \S\,\ref{sec:modeling-pheno}. For clarity, the data are binned in excess of the signal-to-noise ratio of 10 per bin which was used for the modeling. Residuals are shown with colored lines (matching the models in the left panel) for FPMA and in black lines for FPMB. The residuals of the {\tt compTT} models in either geometry are indistinguishable from those in the bottom panel, and are therefore not shown here. }
\label{fig:fig1}
\end{center}
\end{figure*} %%%%%%%%%%%%%%%%%%%%%%%%%%%%%%%%%%%%%%%%%%%%%%%%%%%%%%%%%%%%%%%%%%%%%%%%%%%%%%%%%%%%
%%%%%%%%%%%%%%%%%%%%%%%%%%%%%%%%%%%%%%%%%%%%%%%%%%%%%%%%%%%%%%%%%%%%%%%%%%%%%%%%%%%%%%%%%%%%%%%%%%

%%%%%%%%%%%%%%%%%%%%%%%%%%%%%%%%%%%%%%%%%%%%%%%%%%%%%%%%%%%%%%%%%%%%%%%%%%%%%%%%%%%%%%%%%%%%%%%%%%
\subsection{Phenomenological Models} %%%%%%%%%%%%%%%%%%%%%%%%%%%%%%%%%%%%%%%%%%%%%%%%%%%%%%%%%%%%%
%%%%%%%%%%%%%%%%%%%%%%%%%%%%%%%%%%%%%%%%%%%%%%%%%%%%%%%%%%%%%%%%%%%%%%%%%%%%%%%%%%%%%%%%%%%%%%%%%%

\label{sec:modeling-pheno}

We start the analysis with a simple absorbed power-law model: {\tt TBabs$\times$zTBabs$\times$pow} in \xspec$\!\!$. The first absorption component ({\tt TBabs}; \citealt{wilms+2000}) represents Galactic absorption fixed to a column density of $N_{\rm H, gal}=8\times10^{20}$~cm$^{-2}$ \citep{kalberla+2005}, while the redshifted component ({\tt zTBabs}) accounts for additional absorption by the host galaxy. The redshift is fixed to $z=0.0085$ \citep{wegner+2003}, and the host column density is a free parameter in the fit. This model fits the data from the science observation very poorly, with a reduced $\chi^2$ ($\chi^2/\nu$, where $\nu$ is the number of degrees of freedom), in excess of 3. The best-fit model for the science observation data and the residuals are shown in Figure~\ref{fig:fig1}, in order to highlight the main features that hint towards more appropriate models.

The residuals in the top right panel of Figure~\ref{fig:fig1} show signatures of a reprocessed (reflected) component: a neutral iron K$\alpha$ emission line (6.4~keV) and a broad Compton hump peaking at 20--30~keV. We therefore replace the continuum of the previous model with a \pexrav component \citep{magdziarz+zdziarski-1995}, and add two Gaussian components (one broad and one unresolved; following \citealt{zoghbi+2013}) to model the line. \pexrav includes both the intrinsic power-law continuum and the reflection of that continuum from optically thick material. We keep the inclination angle fixed at the default value ($\cos i = 0.45$, $i\approx60^{\circ}$) and leave chemical abundances fixed at Solar values. For the initial fit, we also keep the energy of the power-law cut-off fixed far above the upper end of the \nustar bandpass at 1000~keV.

This model fits the science observation data much better than the previous one ($\chi^2/\nu$=1513/1124=1.35). The best-fit photon index and absorption column density are $\Gamma=2.00\pm0.01$ and $N_{\rm H}=(2.5\pm0.2)\times10^{22}$~cm$^{-2}$, respectively. The broad Gaussian line component ($\sigma_1=0.35\pm0.03$~keV) is best fitted at a slightly higher energy than the neutral iron K$\alpha$ line: $E_1=6.7\pm0.2$~keV. Fitting for the energy of the narrow line component does not improve the best fit significantly ($\Delta\chi^2/\Delta\nu=-1/-1$), so we leave it fixed at 6.4~keV. The reflection is found to be strong, with a relative normalization $R=0.93\pm0.04$, but clearly insufficient to account for all the curvature present in the hard X-ray spectrum -- as indicated by the residuals of the best fit displayed in the middle right panel of Figure~\ref{fig:fig1}.

Letting the cut-off energy vary in the optimization results in a significant improvement of the best fit: $\chi^2/\nu$=1163/1124=1.03 ($\Delta\chi^2=-349$ for one additional free parameter). This verifies that a cut-off at $E_{\rm cut}\approx115$~keV is robustly detected within the \nustar band. The best fit column density is $N_{\rm H}=(1.1\pm0.2)\times10^{22}$~cm$^{-2}$, which is consistent with the much more precise measurement, $N_{\rm H}=(1.32\pm0.02)\times10^{22}$~cm$^{-2}$, from the joint modeling of the simultaneous \nustar and \suzaku data (Zoghbi {\em et al.}, {\em in prep.}) Freezing $N_{\rm H}$ to $1.32\times10^{22}$~cm$^{-2}$ results in $\Delta\chi^2/\Delta\nu=+3/+1$. For consistency with our work on the joint dataset, we keep $N_{\rm H}$ fixed hereafter. The best-fit parameters of the \pexrav component are $\Gamma=1.85\pm0.01$, $R=0.87\pm0.04$ and $E_{\rm cut}=116_{-5}^{+6}$~keV. The broad iron line is best fitted with $E_1=6.43\pm0.05$~keV and $\sigma_1=0.46\pm0.06$~keV. The model curve and the residuals are plotted in comparison to the previous ones in Figure~\ref{fig:fig1}.

The final form of our phenomenological model is {\tt TBabs$\times$zTBabs$\times$(zgauss[$\times2$]+pexrav)}. Applying this model to the data from the calibration observation, we find that most of the best-fit spectral parameters are consistent with those of the longer science observation (the exception being $R$), albeit less well constrained due to lower photon statistics. The best-fit parameters and their 90\% confidence intervals are given in Table~\ref{tab:model_parameters} for both observations. The flux was $(12\pm1)$\% lower in the 2--10~keV band during the calibration observation, but the two observations can be modeled self-consistently with just the normalization of the primary continuum and the relative reflection normalization changing significantly between the observations. Although we explored other models suggested in the literature, we find that neither adding a second reflection component, nor replacing the \pexrav and the line components with \pexmon (linking those components self-consistently; \citealt{nandra+2007}), nor modeling the broad iron line with a relativistic broadening model, reaches lower $\chi^2/\nu$. More importantly, those alternative models confirm the measurement of $E_{\rm cut}$ to be robust and, in the worst case, marginally consistent with the 90\% confidence interval based on the phenomenological model presented here. This is discussed further in \S\,\ref{sec:discussion-robustness}.

%%%%%%%%%%%%%%%%%%%%%%%%%%%%%%%%%%%%%%%%%%%%%%%%%%%%%%%%%%%%%%%%%%%%%%%%%%%%%%%%%%%%%%%%%%%%%%%%%%
\subsection{Physical Models of the Corona} %%%%%%%%%%%%%%%%%%%%%%%%%%%%%%%%%%%%%%%%%%%%%%%%%%%%%%%
%%%%%%%%%%%%%%%%%%%%%%%%%%%%%%%%%%%%%%%%%%%%%%%%%%%%%%%%%%%%%%%%%%%%%%%%%%%%%%%%%%%%%%%%%%%%%%%%%%

\label{sec:modeling-phys}

In the previous section we established that the coronal continuum can be approximated as a power law with an exponential cut-off at high energies. More physical models (such as the {\tt compTT} model of \citealt{titarchuk-1994}) assume a geometry for the corona and allow for determination of its physical parameters from the data. In such models, low-energy ($\sim$UV) photons from the accretion disk are Compton-scattered by hot electrons in the plasma. The spatial distribution of the coronal plasma can be approximated with simple geometrical shapes, such as a sphere centered on the black hole, or a slab covering the surface of the accretion disk. In \xspec terminology, we replace the \pexrav continuum with a {\tt refl(compTT)} component: {\tt compTT} models the intrinsic coronal continuum for either a slab (disk-like) or a spherical geometry, and {\tt refl} convolves it with reflection features. We fix the thermal photon temperature to 30~eV, which is appropriate for an AGN accretion disk and does not influence the output spectrum much. We leave the reflector inclination fixed at $\cos i=0.45$ and iron abundance fixed at the Solar value.

\begin{figure}[t!] %%%%%%%%%%%%%%%%%%%%%%%%%%%%%%%%%%%%%%%%%%%%%%%%%%%%%%%%%%%%%%%%%%%%%%%%%%%%%%%
\begin{center}
\includegraphics[width=\columnwidth]{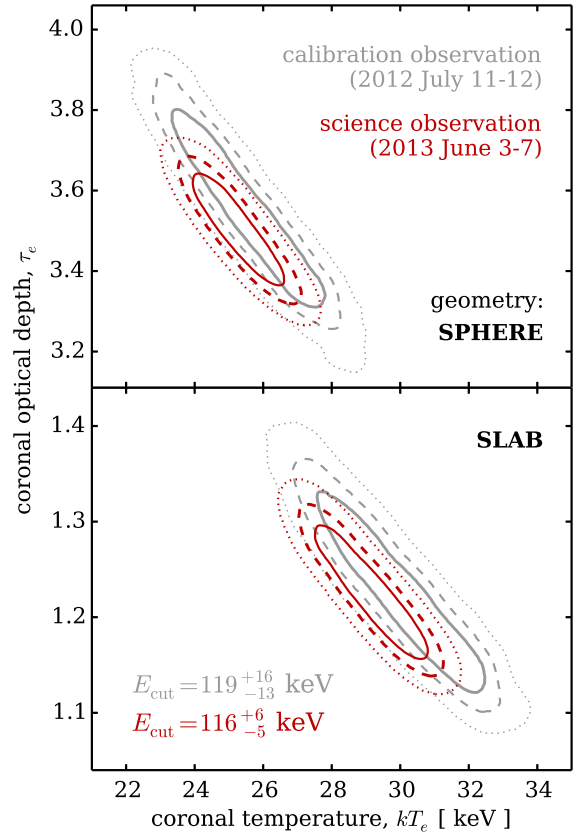}
\caption{ Marginal probability distributions for parameters $\tau_e$ and $kT_e$ of the {\tt compTT} model in the spherical geometry (top panel) and slab geometry (bottom panel). The distributions are derived from MCMC chains computed with \xspec$\!\!$ and normalized separately. The red (grey) contours are based on fits to the science (calibration) observation data, marking enclosed probability of 68, 90 and 99\% with the solid, dashed and dotted lines. }
\label{fig:fig2}
\end{center}
\end{figure}

We find that both geometries can provide a good description of the science observation data: the best-fit $\chi^2$ is 1163 for the slab model, and 1161 for the spherical model, both with 1124 degrees of freedom. In either geometry the coronal temperature ($kT_e$) and the optical depth ($\tau_e$) are very well constrained and strongly correlated, as shown in Figure~\ref{fig:fig2}. In the case of a slab geometry we find $kT_e=29\pm2$~keV and $\tau_e=1.23\pm0.08$, while for the spherical one the best fit is found for $kT_e=25\pm2$~keV and $\tau_e=3.5\pm0.2$. All other parameters are found to be consistent with values determined from the simpler phenomenological models. We find qualitatively and quantitatively similar results for the calibration observation data. Finally, we also verify that consistent results are obtained with a more elaborate coronal model, {\tt compPS} \citep{poutanen+svensson-1996}. While the best-fit parameters may not agree with the {\tt compTT} values within the uncertainties in all cases, the results are qualitatively the same. A complete summary of the best-fit parameters is given in Table~\ref{tab:model_parameters}.

%%%%%%%%%%%%%%%%%%%%%%%%%%%%%%%%%%%%%%%%%%%%%%%%%%%%%%%%%%%%%%%%%%%%%%%%%%%%%%%%%%%%%%%%%%%%%%%%%%%
\begin{deluxetable}{rlcc} %%%%%%%%%%%%%%%%%%%%%%%%%%%%%%%%%%%%%%%%%%%%%%%%%%%%%%%%%%%%%%%%%%%%%%%%%
\tabletypesize{\scriptsize}
\tablewidth{0cm}
\tablecolumns{4}

\tablecaption{ Summary of best-fit model parameters. Uncertainties listed here are 90\% confidence intervals derived from MCMC chains. \label{tab:model_parameters} }

\tablehead{
  \multicolumn{2}{r}{\bf observation \ \ } &
  \colhead{\bf science} &
  \colhead{\bf calibration} \\
  \multicolumn{2}{r}{start--stop date \ \ } &
  \colhead{2013 June 3--7} &
  \colhead{2012 July 11--12} \\
  \multicolumn{2}{r}{$F(2-10~\mbox{keV})$\,\tablenotemark{a} \ \ } &
  \colhead{$10.49\pm0.02$} &
  \colhead{$9.13\pm0.03$} \\
  \multicolumn{2}{r}{$L(2-10~\mbox{keV})$\,\tablenotemark{b} \ \ } &
  \colhead{$1.781\pm0.003$} &
  \colhead{$1.530\pm0.005$} \\
  \multicolumn{2}{r}{d.o.f. ($\nu$) \ \ } &
  \colhead{1124} &
  \colhead{703}
}

\startdata %%%%%%%%%%%%%%%%%%%%%%%%%%%%%%%%%%%%%%%%%%%%%%%%%%%%%%%%%%%%%%%%%%%%%%%%%%%%%%%%%%%%%%

\cutinhead{ \bf independent of the continuum model } %%%%%%%%%%%%%%%%%%%%%%%%%%%%%%%%%%%%%%%%%%%%
$C_{\rm \tiny{FPMB}}$\,\tablenotemark{c}			& & $1.032\pm0.002$ & $1.045\pm0.005$ \\
$E_{\rm line\,1}$									& [ keV ] & $6.43\pm0.05$ & $6.5_{-0.1}^{+0.2}$ \\
$\sigma_{\rm line\,1}$								& [ keV ] & $0.46\pm0.06$ & $0.5\pm0.2$ \\
EW$_{\rm line\,1}$									& [ eV ] & $80\pm10$ & $80\pm20$ \\
EW$_{\rm line\,2}$\									& [ eV ] & $40\pm10$ & $50\pm20$ \\

\cutinhead{ \bf phenomenological continuum model: {\tt pexrav} } %%%%%%%%%%%%%%%%%%%%%%%%%%%%%%%%

$\chi^2$											& & 1163 & 687 \\
$\Gamma$											& & $1.85\pm0.01$ & $1.83\pm0.02$ \\
$R$													& & $0.87\pm0.04$ & $1.1\pm0.1$ \\
$E_{\rm cut}$ 										& [ keV ] & $116_{-5}^{+6}$ & $119_{-13}^{+16}$ \\

\cutinhead{ \bf Comptonized continuum model: {\tt refl(compTT)} } %%%%%%%%%%%%%%%%%%%%%%%%%%%%%%%
\multicolumn{4}{c}{assumed corona geometry: slab} \\ %%%%%%%%%%%%%%%%%%%%%%%%%%%%%%%%%%%%%%%%%%%%
\cline{1-4}

$\chi^2$											& & 1163 & 688 \\
$R$													& & $0.84\pm0.04$ & $1.1\pm0.1$ \\
$kT_e$												& [ keV ] & $29\pm2$ & $30\pm3$ \\
$\tau_e$											& & $1.23\pm0.08$ & $1.2\pm0.1$ \\

\cline{1-4}
\multicolumn{4}{c}{assumed corona geometry: sphere} \\ %%%%%%%%%%%%%%%%%%%%%%%%%%%%%%%%%%%%%%%%%%
\cline{1-4}

$\chi^2$											& & 1161 & 688 \\
$R$													& & $0.82\pm0.04$ & $1.0\pm0.1$ \\
$kT_e$												& [ keV ] & $25\pm2$ & $26\pm3$ \\
$\tau_e$											& & $3.5\pm0.2$ & $3.5\pm0.3$ \\

\cutinhead{ \bf Comptonized continuum model: {\tt compPS} } %%%%%%%%%%%%%%%%%%%%%%%%%%%%%%%%%%%%%
\multicolumn{4}{c}{assumed corona geometry: slab} \\ %%%%%%%%%%%%%%%%%%%%%%%%%%%%%%%%%%%%%%%%%%%%
\cline{1-4}

$\chi^2$											& & 1159 & 690 \\
$R$													& & $0.65\pm0.05$ & $0.83\pm0.09$ \\
$kT_e$												& [ keV ] & $26\pm2$ & $26\pm3$ \\
$\tau_e$											& & $2.2\pm0.1$ & $2.2\pm0.2$ \\

\cline{1-4}
\multicolumn{4}{c}{assumed corona geometry: sphere} \\ %%%%%%%%%%%%%%%%%%%%%%%%%%%%%%%%%%%%%%%%%%
\cline{1-4}

$\chi^2$											& & 1161 & 691 \\
$R$													& & $0.69\pm0.04$ & $0.89\pm0.08$ \\
$kT_e$												& [ keV ] & $25\pm2$ & $25\pm3$ \\
$\tau_e$											& & $3.2\pm0.2$ & $3.3\pm0.3$ \\

\enddata %%%%%%%%%%%%%%%%%%%%%%%%%%%%%%%%%%%%%%%%%%%%%%%%%%%%%%%%%%%%%%%%%%%%%%%%%%%%%%%%%%%%%%%%

\tablenotetext{a}{Flux in the 2--10~keV band in units of $10^{-11}$~erg\,s$^{-1}$\,cm$^{-2}$, calculated from the best-fit phenomenological model. Note that this is an extrapolation down to 2~keV, but we provide it here for comparison with the literature.}
\tablenotetext{b}{Intrinsic continuum luminosity (de-absorbed and excluding reflection components) in the 2--10~keV band in units of $10^{43}$~erg\,s$^{-1}$, calculated from the best-fit phenomenological model. Note that this is an extrapolation down to 2~keV.}
\tablenotetext{c}{Cross-normalization constant for \nustar module FPMB, assuming $C_{\rm \tiny{FPMA}}=1$.}

\end{deluxetable} %%%%%%%%%%%%%%%%%%%%%%%%%%%%%%%%%%%%%%%%%%%%%%%%%%%%%%%%%%%%%%%%%%%%%%%%%%%%%%%%
%%%%%%%%%%%%%%%%%%%%%%%%%%%%%%%%%%%%%%%%%%%%%%%%%%%%%%%%%%%%%%%%%%%%%%%%%%%%%%%%%%%%%%%%%%%%%%%%%%

%%%%%%%%%%%%%%%%%%%%%%%%%%%%%%%%%%%%%%%%%%%%%%%%%%%%%%%%%%%%%%%%%%%%%%%%%%%%%%%%%%%%%%%%%%%%%%%%%%
 %%%%%%%%%%%%%%%%%%%%%%%%%%%%%%%%%%%%%%%%%%%%%%%%%%%%%%%%%%%%%%%%%%%%%%%%%%%%%%%%%%%%%
%%%%%%%%%%%%%%%%%%%%%%%%%%%%%%%%%%%%%%%%%%%%%%%%%%%%%%%%%%%%%%%%%%%%%%%%%%%%%%%%%%%%%%%%%%%%%%%%%%

\vspace{1cm}

%%%%%%%%%%%%%%%%%%%%%%%%%%%%%%%%%%%%%%%%%%%%%%%%%%%%%%%%%%%%%%%%%%%%%%%%%%%%%%%%%%%%%%%%%%%%%%%%%%
%%%%%%%%%%%%%%%%%%%%%%%%%%%%%%%%%%%%%%%%%%%%%%%%%%%%%%%%%%%%%%%%%%%%%%%%%%%%%%%%%%%%%%%%%%%%%%%%%%
\section{Discussion} %%%%%%%%%%%%%%%%%%%%%%%%%%%%%%%%%%%%%%%%%%%%%%%%%%%%%%%%%%%%%%%%%%%%%%%%%%%%%
%%%%%%%%%%%%%%%%%%%%%%%%%%%%%%%%%%%%%%%%%%%%%%%%%%%%%%%%%%%%%%%%%%%%%%%%%%%%%%%%%%%%%%%%%%%%%%%%%%
%%%%%%%%%%%%%%%%%%%%%%%%%%%%%%%%%%%%%%%%%%%%%%%%%%%%%%%%%%%%%%%%%%%%%%%%%%%%%%%%%%%%%%%%%%%%%%%%%%

\label{sec:discussion}

%%%%%%%%%%%%%%%%%%%%%%%%%%%%%%%%%%%%%%%%%%%%%%%%%%%%%%%%%%%%%%%%%%%%%%%%%%%%%%%%%%%%%%%%%%%%%%%%%%
\subsection{The Hard X-ray Spectrum and Its Variability} %%%%%%%%%%%%%%%%%%%%%%%%%%%%%%%%%%%%%%%%%
%%%%%%%%%%%%%%%%%%%%%%%%%%%%%%%%%%%%%%%%%%%%%%%%%%%%%%%%%%%%%%%%%%%%%%%%%%%%%%%%%%%%%%%%%%%%%%%%%%

\label{sec:discussion-model}

Our spectral modeling results are generally consistent with previous findings, and confirm that the X-ray spectrum of \mcg resembles that of a classical Compton-thin Seyfert~2 nucleus (e.g., \citealt{walton+2013}). The high-energy cut-off has been previously measured in \mcg with the \bepposax$\!\!$, \integral and \swift$\!\!$ hard X-ray instruments: $147_{-40}^{+70}$~keV \citep{perola+2002}, $190_{-60}^{+110}$~keV \citep{dadina-2007}, $85_{-20}^{+35}$~keV \citep{molina+2013}. \citet{beckmann+2008} claimed that the cut-off energy is variable within the 50$\sim$100~keV range, but did not highlight any clear trends. It is important to stress that these inferences required assumptions about the photon index and reflection normalization in most cases, while we determine these spectral parameters directly from the data. The phenomenological model presented in \S\,\ref{sec:modeling-pheno} is the simplest model accounting for the key spectral features observed in the \nustar bandpass: the iron lines, the Compton hump, and the high-energy cut-off. We emphasize that it should not be taken too literally, as we exploit its simplicity only to highlight the precision of the $E_{\rm cut}$ measurement and the issues that it raises. 

More complicated models are clearly needed to fully explain the high-quality soft X-ray observations (e.g., \citealt{reeves+2007,zoghbi+2013}; also Zoghbi {\em et al.}, {\em in prep.}). Although \nustar does not have sufficient spectral resolution to resolve details in the iron line complex, we compute equivalent widths of the two Gaussian components used in our modeling ($80\pm10$~eV for the broad and $40\pm10$~eV for the narrow component; see Table~\ref{tab:model_parameters}) and find that they are consistent with the highest-quality soft X-ray data. We also test a two-component reflection model, in which the distant reflection is separated from the relativistically broadened and partially ionized reflection off the inner accretion disk. For the disk reflection component we use {\tt reflionx\_hc}---an updated version of {\tt reflionx} \citep{ross+fabian-2005} with a variable $E_{\rm cut}$---and relativistic broadening modeled by a convolution with the \xspec model {\tt kdblur}. We find that the \nustar data are not sensitive to the accretion disk parameters as long as its ionization is low ($\xi\lesssim50$~erg\,s\,cm$^{-1}$), which is suggested by the best fit. Although the exact best-fit $E_{\rm cut}$ depends on the nuisance parameters, in all cases it is found to be marginally consistent (at the 90\% confidence level) with $E_{\rm cut}=116_{-5}^{+6}$~keV.

The variability on the $\lesssim$1-ks timescale has been shown to be due to reverberation of the primary continuum on the inner accretion disk \citep{zoghbi+2014}. Evaluation of the spectral variability between the two \nustar observations (approximately one year apart) is limited by the possible variability of the absorbing column. Our analysis of the joint \nustar and \suzaku dataset from 2013 gives a relatively low absorption column density compared to the average taken from the literature ($1.32\times10^{22}$~cm$^{-2}$ compared to $\approx1.6\times10^{22}$~cm$^{-2}$, excluding the Galactic contribution), which might or might not have persisted since the calibration observation in 2012. In our modeling, summarized in Table~\ref{tab:model_parameters}, we assume the same absorption column for both observations. If we instead adopt the long-term average column for the calibration observation,\footnote{Due to the lack of coverage below 3~keV, \nustar alone cannot constrain strongly column densities as low as $1\times10^{22}$~cm$^{-2}$. With $N_{\rm H}$ left free to vary, the best fit for the calibration observation is found for $(1.3\pm0.4)\times10^{22}$~cm$^{-2}$.} we find a cut-off at $\approx$130~keV, which is only marginally different from the science observation. With no soft X-ray coverage for the calibration observation, the claim that $E_{\rm cut}$ is variable is therefore not statistically significant.

With the column density kept constant, only the absolute flux and the relative reflection normalization ($R$) seem to have changed significantly. If we separate the reflection from the coronal continuum,\footnote{This is achieved by setting the {\tt pexrav} component to produce only the reflection continuum (formally, $R<0$ in \xspec$\!\!$) and adding a separate cut-off power law continuum component, where the photon index and the cut-off energy are shared by both components and their normalizations are free to vary independently.} we find that the normalization of the former does not change significantly between the two observations and conclude that the change is due to the relative increase of the coronal continuum flux. The flux of the broad iron line component is constant between the observations within the 90\% confidence interval. The spectral variability may be due to the time delay between variations in the coronal continuum and its reflection by the distant torus. Alternatively, an effective change in $R$ may be due to a long-term physical change in the coronal geometry, such as its height above the accretion disk, or in the innermost region of the accretion disk itself.

%%%%%%%%%%%%%%%%%%%%%%%%%%%%%%%%%%%%%%%%%%%%%%%%%%%%%%%%%%%%%%%%%%%%%%%%%%%%%%%%%%%%%%%%%%%%%%%%%%
\subsection{Robustness of the Cut-off Measurement} %%%%%%%%%%%%%%%%%%%%%%%%%%%%%%%%%%%%%%%%%%%%%%%
%%%%%%%%%%%%%%%%%%%%%%%%%%%%%%%%%%%%%%%%%%%%%%%%%%%%%%%%%%%%%%%%%%%%%%%%%%%%%%%%%%%%%%%%%%%%%%%%%%

\label{sec:discussion-robustness}

As demonstrated in \S\,\ref{sec:modeling-pheno} (see Fig.~\ref{fig:fig1}), a high-energy cut-off is clearly required by the \nustar data. Even though the cut-off energy ($E_{\rm cut}$) is above the upper end of the \nustar bandpass, strong curvature is apparent below 79~keV and allows for determination of $E_{\rm cut}$ to $\lesssim5$\% (statistical uncertainty; 90\% confidence). The best-fit value of $E_{\rm cut}$, however, depends on the assumptions that go into the simple model we fit to the data. One example is the inclination: if left free to vary in optimization, the best fit tends to $i\approx80^{\circ}$ and $E_{\rm cut}\approx130$~keV, whereas adopting a value from the recent literature\footnote{Note that \citet{weaver+1997} and \citet{mattson+weaver-2004} advocated a nearly face-on inclination ($i\approx80^{\circ}$); however, more recent data do not favor that value.} ($i\approx45^{\circ}$; \citealt{braito+2007,reeves+2007,zoghbi+2013}) leads to $E_{\rm cut}\approx110$~keV. Likewise, if we leave the iron abundance to vary freely, the best fit is found for $A_{\rm Fe}=0.9\pm0.2$ -- this is consistent with our assumption of $A_{\rm Fe}=1$, but implies $E_{\rm cut}=122$~keV, which is at the upper end of the 90\% confidence interval found in \S\,\ref{sec:modeling-pheno}.

A two-component reflection model leads to best-fit $E_{\rm cut}$ between 110 and 124~keV, depending on different assumptions. The typical statistical uncertainty on the best-fit $E_{\rm cut}$ in any particular fit to the science observation data is approximately 7~keV (20--30~keV for the calibration observation), with the iron abundance left free to vary and the ionization and the relativistic broadening parameters fixed close to values found in previous work (e.g., \citealt{zoghbi+2013}). We emphasize, however, that the systematics introduced by assuming a particular model are comparable to the statistical uncertainties in the case of the science observation of \mcg$\!\!$, and are therefore important to consider. With the full flexibility in the shape of the complex reflection continuum, the \nustar data robustly constrain $E_{\rm cut}$ to the slightly broader 105--130~keV interval, skewed towards the lower end and centered around 115~keV (when marginalized over different assumptions).

For high-quality data systematic uncertainty comparable to statistical uncertainty may also arise from arbitrary choices of the source and background extraction regions, and the choice of binning. For the $E_{\rm cut}$ measurement presented in this paper, we have verified that different choices give results consistent with those discussed above. Systematics are clearly less of an issue with lower-quality data, as demonstrated by the calibration observation data presented here: in that case the constraints on spectral parameters are weakened, and the systematic uncertainty gets absorbed in the statistical uncertainty. This has been the case for the majority of the similar measurements on other AGN published so far, including the recent ones based on the \nustar data \citep{brenneman+2014a,marinucci+2014,ballantyne+2014}. As in the case of IC~4329a \citep{brenneman+2014b}, additional constraints come from joint analyses of simultaneous soft and hard X-ray datasets, leading to further improvement in constraining $E_{\rm cut}$.

%%%%%%%%%%%%%%%%%%%%%%%%%%%%%%%%%%%%%%%%%%%%%%%%%%%%%%%%%%%%%%%%%%%%%%%%%%%%%%%%%%%%%%%%%%%%%%%%%%
\subsection{Towards a Physical Model of the AGN Corona} %%%%%%%%%%%%%%%%%%%%%%%%%%%%%%%%%%%%%%%%%%
%%%%%%%%%%%%%%%%%%%%%%%%%%%%%%%%%%%%%%%%%%%%%%%%%%%%%%%%%%%%%%%%%%%%%%%%%%%%%%%%%%%%%%%%%%%%%%%%%%

\label{sec:discussion-physics}

The high-energy cut-offs have been measured with a relative uncertainty of $\gtrsim$30\% for a relatively small sample of bright nearby AGN; most of the AGN observed with previous generation of hard X-ray instruments provide lower limits on this parameter (e.g., \citealt{dadina-2007,malizia+2014}). Using the \nustar data, the cut-off energies have recently been measured for IC~4329a ($E_{\rm cut}=184\pm14$~keV; \citealt{brenneman+2014b}), SWIFT~J2127.4$+$5654 ($E_{\rm cut}=108_{-10}^{+11}$~keV; \citealt{marinucci+2014}) and 3C~382 ($E_{\rm cut}=214_{-63}^{+147}$ and $>190$~keV in two distinct spectral states; \citealt{ballantyne+2014}). The high quality of the \nustar spectra in the hard X-ray band up to 79~keV enable reliable independent measurements of the physical parameters of the corona: its temperature, $kT_e$, and optical depth, $\tau_e$. In this paper we present the most precise measurement thus far, with relative uncertainty of 5\% (at the 90\% confidence level), although, as discussed in \S\,\ref{sec:discussion-robustness}, the exact value depends somewhat on nuisance parameters.

We find that, even though the physical parameters such as $kT_e$ and $\tau_e$ are very well constrained by the data, it is still impossible to formally distinguish the geometry. The slab (disk-like) and the spherical geometries, as parametrized by the {\tt compTT} and {\tt compPS} models used here, both describe the \mcg spectrum equally well. We note that a similar result was found in observations of the Seyfert~1.2 IC~4329a and the narrow-line Seyfert~1 SWIFT~J2127.4$+$5654 with \nustar \citep{brenneman+2014a,brenneman+2014b,marinucci+2014}. Both of these AGN and \mcg are radio-quiet, however, they differ in other properties. With a mass of the super-massive black hole of $\sim5\times10^7$\,$M_{\odot}$ \citep{wandel+mushotzky-1986}, the mean intrinsic 2--10~keV luminosity of $1.66\times10^{43}$~erg\,s$^{-1}$ (see Table~\ref{tab:model_parameters}) and a bolometric correction from \citet{marconi+2004}, \mcg is accreting at approximately 5\% of the Eddington rate. This is almost an order of magnitude less than the key other two AGN. Interestingly, SWIFT~J2127.4$+$5654 has the lowest black hole mass and the lowest cut-off, followed by \mcg in the middle, and IC~4329a with highest mass and cut-off energy. In a number of other AGN, a stringent lower limit on the cut-off energy was placed using the \nustar data, indicating a generally higher coronal temperature and lower optical depths, e.g., $E_{\rm cut}>190$~keV in 3C~382 \citep{ballantyne+2014} and in Ark~120~\citep{matt+2014}, and $E_{\rm cut}>210$~keV in NGC~2110~\citep{marinucci+2015}. Using long-term averaged data from \integral$\!\!$, \citet{malizia+2014} constrained cut-off energies for 26 AGN in the range between 50 and 200~keV, some of which have been or will be observed with \nustar$\!\!$. With more high-quality measurements in the near future, covering a wide range of physical properties, it will be possible to directly probe the physics of the AGN corona. In order to distinguish the fine differences due to the coronal geometry, longer observations of sources with a weaker reflection continuum will be needed.

The difference between the optical depth in the two geometries tested here is partially due to the different geometrical definition: whereas in the spherical case it is taken in the radial direction, in the case of slab geometry it is taken vertically, creating a natural offset by a factor of $\cos i$. For $\cos i=0.45$ used here, the radial optical depth for the slab geometry becomes almost equal to the one of the spherical corona. The important result we point out in this paper is that the $E_{\rm cut}<$200~keV measurements with \nustar pressure the theoretical models towards the high-$\tau_e$ regime where their validity falls off. The approximations used in the {\tt compPS} model hold only for low optical depth and the formal limits are $\tau_e<1.5$ for the slab, and $\tau_e<3$ for the sphere geometry \citep{poutanen+svensson-1996}. The limits of the simpler {\tt compTT} model are even more stringent, although good agreement is found between the analytical model and Monte Carlo simulations in the $\tau_e\sim1$ regime \citep{titarchuk-1994}. It is therefore not surprising that the best-fit optical depth in the two models differs somewhat. If the high optical depth derived from the currently available models can be interpreted directly, our results imply that the corona must be inhomogenuous. Spectral features and variabilty signatures of reflection from the inner accretion disk are clearly detectable in \mcg \citep{zoghbi+2014} and therefore the corona, which covers the disk in either geometry, cannot be completely opaque. Homogeneity is one of the assumptions of the coronal models used here, hence pressing against their limits may be indicative of that assumption not being satisfied. Alternatively, our result may simply indicate a geometry different from the ones assumed in this work. In either case, we are drawn to the conclusion that new models are needed in order to better understand the physical implications of our result.

%%%%%%%%%%%%%%%%%%%%%%%%%%%%%%%%%%%%%%%%%%%%%%%%%%%%%%%%%%%%%%%%%%%%%%%%%%%%%%%%%%%%%%%%%%%%%%%%%%
%%%%%%%%%%%%%%%%%%%%%%%%%%%%%%%%%%%%%%%%%%%%%%%%%%%%%%%%%%%%%%%%%%%%%%%%%%%%%%%%%%%%%%%%%%%%%%%%%%
\section{Summary and Conclusion} %%%%%%%%%%%%%%%%%%%%%%%%%%%%%%%%%%%%%%%%%%%%%%%%%%%%%%%%%%%%%%%%%
%%%%%%%%%%%%%%%%%%%%%%%%%%%%%%%%%%%%%%%%%%%%%%%%%%%%%%%%%%%%%%%%%%%%%%%%%%%%%%%%%%%%%%%%%%%%%%%%%%
%%%%%%%%%%%%%%%%%%%%%%%%%%%%%%%%%%%%%%%%%%%%%%%%%%%%%%%%%%%%%%%%%%%%%%%%%%%%%%%%%%%%%%%%%%%%%%%%%%

\label{sec:summary}

In this paper we focus on modeling the hard X-ray spectrum of \mcg in order to constrain models of the AGN corona. We first robustly establish that a cut-off is present in the spectrum at $116_{-5}^{+6}$~keV (statistical uncertainty; 90\% confidence), despite the non-negligible reflection component contributing to curvature of the hard X-ray spectrum. The ability to dissentangle a $\lesssim200$~keV cut-off from the reflection continuum is essentially unique to \nustar$\!\!$. Modeling the spectrum with physical models, we find that both slab and spherical geometries of the corona provide equally good fits to the data, albeit for different physical parameters. Assuming a simple coronal model ({\tt compTT}), we find the kinetic temperature of electrons in the corona and its optical depth, $kT_e$ and $\tau_e$, to be $29\pm2$ ($25\pm2$)~keV and $1.23\pm0.08$ ($3.5\pm0.2$) for the slab (spherical) geometry. Similar results are found for a different, less approximate model ({\tt compPS}). It is important to note that in all cases the data push the models towards high-$\tau_e$ values, where their validity drops off. The relative statistical uncertainty of 5\% (quoted here as a 90\% confidence interval) has never been achieved before and we show that the new level of precision enabled by \nustar requires careful consideration of possible systematic uncertainties arising from simplifying assumptions. With further measurements at comparable precision for AGN with a wide range of properties, and the extension of Comptonization models towards the high-opacity regime, it should be possible to construct a clearer physical picture of the AGN corona in the near future.

%%%%%%%%%%%%%%%%%%%%%%%%%%%%%%%%%%%%%%%%%%%%%%%%%%%%%%%%%%%%%%%%%%%%%%%%%%%%%%%%%%%%%%%%%%%%%%%%%%
\acknowledgements %%%%%%%%%%%%%%%%%%%%%%%%%%%%%%%%%%%%%%%%%%%%%%%%%%%%%%%%%%%%%%%%%%%%%%%%%%%%%%%%
%%%%%%%%%%%%%%%%%%%%%%%%%%%%%%%%%%%%%%%%%%%%%%%%%%%%%%%%%%%%%%%%%%%%%%%%%%%%%%%%%%%%%%%%%%%%%%%%%%

M.\,B. acknowledges support from the International Fulbright Science and Technology Award. A.\,M. and G.\,M. acknowledge financial support from Italian Space Agency under grant ASI/INAF I/037/12/0-011/13 and from the European Union Seventh Framework Programme (FP7/2007-2013) under grant agreement n.\,312789. C.\,S.\,R. thanks NASA for support under ADAP grant NNX14AF86G. This work was supported under NASA Contract No.~NNG08FD60C, and made use of data from the \nustar mission, a project led by the California Institute of Technology, managed by the Jet Propulsion Laboratory, and funded by the National Aeronautics and Space Administration. We thank the \nustar Operations, Software and Calibration teams for support with the execution and analysis of these observations. This research has made use of the \nustar Data Analysis Software (NuSTARDAS) jointly developed by the ASI Science Data Center (ASDC, Italy) and the California Institute of Technology (USA). This research has made use of NASA's Astrophysics Data System.

{\it Facilities:} \nustar

%%%%%%%%%%%%%%%%%%%%%%%%%%%%%%%%%%%%%%%%%%%%%%%%%%%%%%%%%%%%%%%%%%%%%%%%%%%%%%%%%%%%%%%%%%%%%%%%%%
{} %%%%%%%%%%%%%%%%%%%%%%%%%%%%%%%%%%%%%%%%%%%%%%%%%%%%%%%%%%%%%%%%%%%%%%%%%%
%%%%%%%%%%%%%%%%%%%%%%%%%%%%%%%%%%%%%%%%%%%%%%%%%%%%%%%%%%%%%%%%%%%%%%%%%%%%%%%%%%%%%%%%%%%%%%%%%%


\begin{thebibliography}{} %%%%%%%%%%%%%%%%%%%%%%%%%%%%%%%%%%%%%%%%%%%%%%%%%%%%%%%%%%%%%%%%%%%%%%%%

\bibitem[Arnaud(1996)]{arnaud+1996} Arnaud, K.\,A. 1996, in ASP Conf. Ser. 101: {\it Astronomical Data Analysis Software and Systems V XSPEC: The First Ten Years}, p. 17
\bibitem[Ballantyne(2014)]{ballantyne-2014} Ballantyne, D.\,R. 2014, MNRAS, 437, 2845
\bibitem[Ballantyne \etal(2014)]{ballantyne+2014} Ballantyne, D.\,R., Bollenbacher, J.\,M., Brenneman, L.\,W., \etal\ 2014, ApJ, 794, 62
\bibitem[Balestra \etal(2004)]{balestra+2004} Balestra, I., Bianchi, S., \& Matt, G. 2004, A\&A, 415, 437 
\bibitem[Bassani \etal(2006)]{bassani+2006} Bassani, L., Molina, M., Malizia, A., \etal\ 2006, ApJ, 636, L65
\bibitem[Baumgartner \etal(2013)]{baumgartner+2013} Baumgartner, W.\,H., Tueller, J., Markwardt, C.\,B., \etal\ 2013, ApJS, 207, 19
\bibitem[Beckmann \etal(2008)]{beckmann+2008} Beckmann, V., Courvoisier, T.\,J.-L., Gehrels, N., \etal\ 2008, A\&A 492, 93 
\bibitem[Braito \etal(2007)]{braito+2007} Braito, V., Reeves, J.\,N., Dewangan, G.\,C., \etal\ 2007, ApJ, 670, 978
\bibitem[Brenneman \etal(2014a)]{brenneman+2014a} Brenneman, L.\,W., Madejski, G., F\"{u}rst, F., \etal\ 2014, ApJ, 781, 83 
\bibitem[Brenneman \etal(2014b)]{brenneman+2014b} Brenneman, L.\,W., Madejski, G., F\"{u}rst, F., \etal\ 2014, ApJ, 788, 61 
\bibitem[Burlon \etal(2011)]{burlon+2011} Burlon, D., Ajello, M., Greiner, J., \etal\ 2011, ApJ, 728, 58 
\bibitem[Dadina(2007)]{dadina-2007} Dadina, M. 2007, A\&A, 461, 1209
\bibitem[Dai \etal(2010)]{dai+2010} Dai, X., Kochanek, C.\,S., Chartas, G., \etal\ 2010, ApJ, 709, 278
\bibitem[George \& Fabian(1991)]{george+fabian-1991} George I.\,M. \& Fabian A.\,C. 1991, MNRAS, 249, 352
\bibitem[Ghisellini \etal(1994)]{ghisellini+1994} Ghisellini, G., Haardt, F., Matt, G. 1994, MNRAS, 267, 743
\bibitem[Haardt \etal(1994)]{haardt+1994} Haardt, F., Maraschi, L., \& Ghisellini, G., 1994, ApJ, 432, L95
\bibitem[Harrison \etal(2013)]{harrison+2013} Harrison, F.~A., \etal\ 2013, ApJ, 770, 103
\bibitem[Kalberla \etal(2005)]{kalberla+2005} Kalberla, P. M., Burton, W. B., Hartmann, D., \etal\ 2005, A\&A, 440, 775 
\bibitem[Madsen \etal(2015)]{madsen+2015} Madsen, K.\,K. \etal\ 2015, ApJ, submitted
\bibitem[Magdziarz \& Zdziarski(1995)]{magdziarz+zdziarski-1995} Magdziarz \& Zdziarski 1995 MNRAS, 273, 837
\bibitem[Malizia \etal(2014)]{malizia+2014} Malizia, A., Molina, M., Bassani, L., \etal\ 2014, ApJL, 782, 25
\bibitem[Marconi \etal(2004)]{marconi+2004} Marconi, A., Risaliti, G., Gilli, R., \etal\ 2004, MNRAS, 351, 169
\bibitem[Marinucci \etal(2014)]{marinucci+2014} Marinucci, A., Matt, G., Kara, E., \etal\ 2014, MNRAS, 440, 2347
\bibitem[Marinucci \etal(2015)]{marinucci+2015} Marinucci, A., Matt, G., Bianchi, S., \etal\ 2015, MNRAS, 447, 160
\bibitem[Matt \etal(2014)]{matt+2014} Matt, G., Marinucci, A., Guainazzi, M., \etal 2014, 439, 3016
\bibitem[Mattson \& Weaver(2004)]{mattson+weaver-2004} Mattson, B.\,J., \& Weaver, K.\,A. 2004, ApJ, 601, 771
\bibitem[Molina \etal(2013)]{molina+2013} Molina, M., Bassani, L., Malizia, A., \etal\ 2013, MNRAS, 433, 1687
\bibitem[Mosquera \etal(2013)]{mosquera+2013} Mosquera, A.\,M., Kochanek, C.\,S., Chen, B. \etal\ 2013, ApJ, 769, 53 
\bibitem[Nandra \etal(2007)]{nandra+2007} Nandra, P., \etal\ 2007, MNRAS, 382, 194
\bibitem[Perola \etal(2002)]{perola+2002} Perola, G.\,C., Matt, G., Cappi, M., \etal\ 2002, A\&A, 389, 802
\bibitem[Perri \etal(2013)]{perri+2013} Perri, M., Puccetti, S., Spagnuolo, N, \etal\ 2013, {\it The NuSTAR Data Analysis Software Guide}, {\tt http://heasarc.gsfc.nasa.gov/ docs/nustar/ analysis/NuSTARDAS\_swguide\_v1.5.pdf}
\bibitem[Poutanen \& Svensson(1996)]{poutanen+svensson-1996} Poutanen, J., Svensson, R., 1996, ApJ, 470, 249
\bibitem[Reeves \etal(2007)]{reeves+2007} Reeves, J.\,N., Awaki, H., Dewangan, G.\,C., \etal 2007, PASJ, 59, 301
\bibitem[Reis \& Miller(2013)]{reis+miller-2013} Reis, R.~C. \& Miller, J.~M., 2013, ApJ, 769, L7
\bibitem[Ross \& Fabian(2005)]{ross+fabian-2005} Ross, R.\,R., Fabian, A.\,C. 2005, MNRAS, 358, 211
\bibitem[Rybicki \& Lightman(1979)]{rybicki+lightman-1979} Rybicki, G. B., \& Lightman, A. P. 1979, Radiative Processes in Astrophysics (John Wiley \& Sons Inc.)
\bibitem[Titarchuk(1994)]{titarchuk-1994} Titarchuk, L. 1994, ApJ, 434, 570
\bibitem[Vasudevan \etal(2013)]{vasudevan+2013} Vasudevan, R.\,V., Brandt, W.\,N., Mushotzky, R.\,F., \etal\ 2013, ApJ, 763, 111
\bibitem[Veron \etal(1980)]{veron+1980} V\'{e}ron, P., Lindblad, P.~O., Zuiderwijk, E.~J., V\'{e}ron, M.~P., \& Adam, G., 1980, A\&A, 87, 245
\bibitem[Walton \etal(2013)]{walton+2013} Walton, D.\,J., Nardini, E., Fabian, A.\,C., \etal\ 2013, MNRAS, 428, 2901
\bibitem[Wang \& Zhang(2007)]{wang+zhang-2007} Wang, J.-M., \& Zhang, E.~P., 2007, ApJ, 660, 1072
\bibitem[Weaver \etal(1997)]{weaver+1997} Weaver, K.\,A., Yaqoob, T., Mushotzky, R.\,F., \etal\ 1997, ApJ, 474, 675
\bibitem[Wegner \etal(2003)]{wegner+2003} Wegner, G., \etal\ 2003, AJ, 126, 2268
\bibitem[Wandel \& Mushotzky(1986)]{wandel+mushotzky-1986} Wandel, A., \& Mushotzky, R.\,F. 1986, ApJ, 306, 61
\bibitem[Wilms \etal(2000)]{wilms+2000} Wilms, J., Allen, A. \& McCray, R. 2000, ApJ, 542, 914
\bibitem[Zdziarski \etal(2000)]{zdziarski+2000} Zdziarski, A. A., Poutanen, J., \& Johnson, W. N. 2000, ApJ, 542, 703
\bibitem[Zoghbi \etal(2013)]{zoghbi+2013} Zoghbi, A., Reynolds, C., Cackett, E.\,M., \etal\ 2013, ApJ, 767, 121
\bibitem[Zoghbi \etal(2014)]{zoghbi+2014} Zoghbi, A., Cackett, E.\,M., Reynolds, C., \etal\ 2014, ApJ, 789, 56

\end{thebibliography}
\end{document}